\documentclass[10pt,twocolumn,letterpaper]{article}

\usepackage{cvpr}              % To produce the CAMERA-READY version
\usepackage[dvipsnames]{xcolor}

\definecolor{cvprblue}{rgb}{0.21,0.49,0.74}
\usepackage[table]{xcolor} 
\usepackage[pagebackref,breaklinks,colorlinks,citecolor=cvprblue]{hyperref}

\def\paperID{312} % *** Enter the Paper ID here
\def\confName{3DV\xspace}
\def\confYear{2026\xspace}

\begin{document}

\hypersetup{
  colorlinks=true,
  citecolor=cvprblue,
  urlcolor=cvprblue,
  linkcolor=cvprblue
}
% \twocolumn[{%
% \renewcommand\twocolumn[1][]{#1}%
% \input{sec/0_title}%
% \maketitle%
% \begin{center}%
%     \input{sec/0_teaser}%
% \end{center}%
% }]

% \begin{title}

\title{TexAvatars: Hybrid Texel-3D Representations for Stable Rigging of Photorealistic Gaussian Head Avatars}

\author{
Jaeseong Lee$^{1*}$ \quad
Junyeong Ahn$^{2*}$ \quad
Taewoong Kang$^{1}$ \quad
Jaegul Choo$^{1}$ \\ \\
$^{1}$KAIST \hspace{0.4cm}$^{2}$Hanyang University \\
{\tt\small \{webmaster, keh0t0, jchoo\}@kaist.ac.kr, hewas1230@hanyang.ac.kr}}

% \author{
% Jaeseong Lee$^{1*}$ \quad
% Junyeong Ahn$^{2}$\thanks{Equal contribution.} \quad
% Taewoong Kang$^{1}$ \quad
% Jaegul Choo$^{1}$ \\ \\
% $^{1}$KAIST \hspace{0.4cm}$^{2}$Hanyang University \\
% {\tt\small \{webmaster, keh0t0, jchoo\}@kaist.ac.kr, hewas1230@hanyang.ac.kr}}

\twocolumn[{%
\renewcommand\twocolumn[1][]{#1}%  <- 재귀 방지
\maketitle%                         <- 제목/저자 출력
\vspace{-2.2em}  % 붙이는 정도(더 붙이고 싶으면 -2.3em 같은 식)

{\centering
\large \href{https://summertight.github.io/TexAvatars}{summertight.github.io/TexAvatars}\par
}

\vspace{1.2em}  % 아래 그림/본문과의 간격

\begin{center}%                     <- float(figure) 쓰지 말 것!
  % sec/0_teaser.tex 안에도 figure/figure* 쓰지 말고,
  % 그냥 \includegraphics 등 "내용"만 두세요.
  % \begin{teaser}

\vspace{-1em}
    \centering
    \scriptsize
    \captionsetup{type=figure}
    \includegraphics[width=\textwidth]{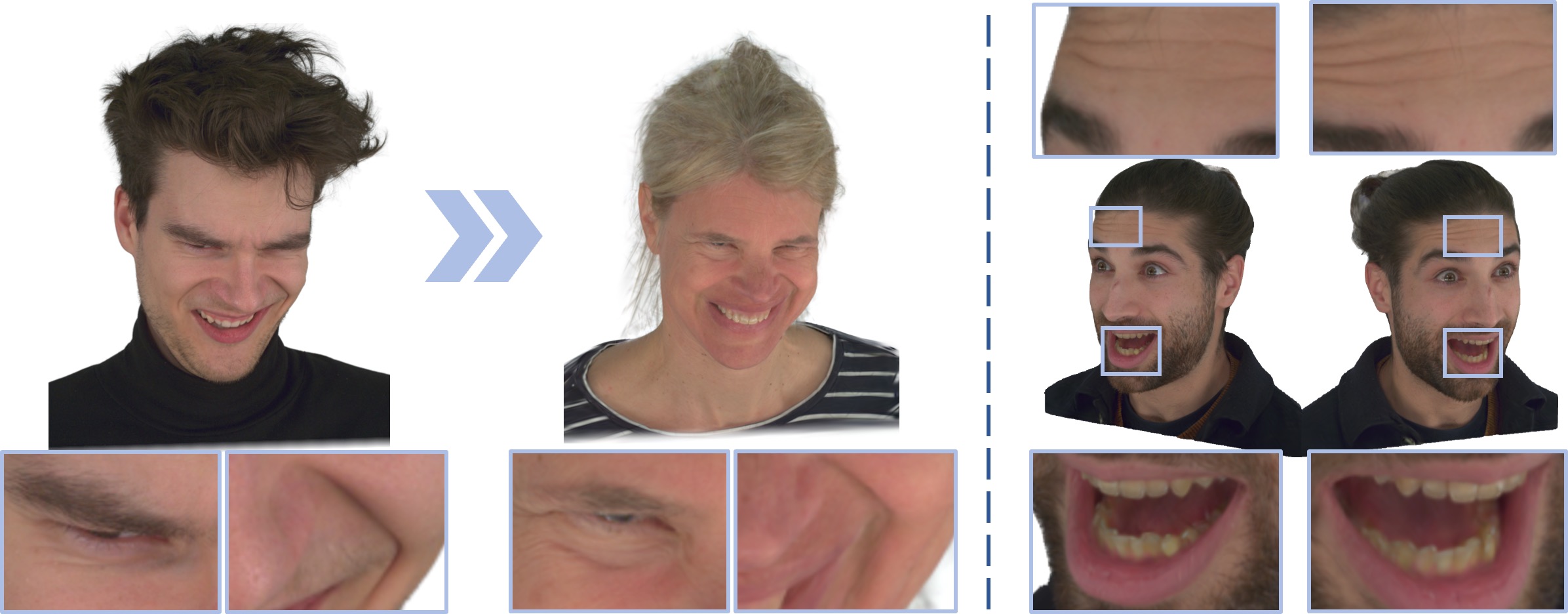}
    \vspace{-1em}
    \caption{We propose a high-fidelity head avatar method that combines analytic rigging with texel-space neural regression. Gaussian attributes are predicted in UV space and lifted to 3D via interpolated deformation fields, enabling photorealistic 3D Gaussian Splatting. Decoupling from fixed triangle bindings yields robust extrapolation to extreme expressions and poses. \textit{Left:} unseen expressions show strong muscle activations. \textit{Right:} wrinkles and teeth remain consistent across mirrored views, highlighting robustness and view consistency.}
\label{fig:teaser}

%   \includegraphics[width=\textwidth]{figures/teaser.jpg}
%   \caption{We present a high-fidelity head avatar method that combines the structural interpretability of analytic rigging with the flexibility of texel space neural regression. Our approach regresses local Gaussian attributes in the UV domain and lifts them to 3D using interpolated mesh-aware deformation fields, enabling photorealistic rendering via 3D Gaussian Splatting. By decoupling deformation from fixed triangle bindings, our model robustly handles extreme expressions and poses. \textit{Left:} reenactment under an extreme unseen expression demonstrates strong extrapolation capability such as muscle activations. \textit{Right:} high-frequency details, wrinkles and teeth, are consistently preserved across mirrored novel views, highlighting view consistency and deformation robustness.
% }

%   \label{fig:teaser}
%
\end{center}%
\par
}]

% \maketitle

% \input{sec/0_teaser} 

\let\thefootnote\relax\footnotetext{$*$ denotes equal contribution.}\par
\begin{abstract}

Constructing drivable and photorealistic 3D head avatars has become a central task in AR/XR, enabling immersive and expressive user experiences. With the emergence of high-fidelity, efficient representations such as 3D Gaussians, recent works have pushed toward ultra-detailed head avatars. Existing approaches typically fall into two categories: rule-based analytic rigging or neural network-based deformation fields. While effective in constrained settings, both approaches often fail to generalize to unseen expressions and poses—particularly in extreme reenactment scenarios. Other methods constrain Gaussians to the global texel space of 3DMMs to reduce rendering complexity. However, these texel-based avatars tend to underutilize the underlying mesh structure. They apply minimal analytic deformation and rely heavily on neural regressors and heuristic regularization in UV space, which weakens geometric consistency and limits extrapolation to complex, out-of-distribution deformations. To address these limitations, we introduce \textbf{TexAvatars}, a hybrid avatar representation that combines the explicit geometric grounding of analytic rigging with the spatial continuity of texel space. Our approach predicts local geometric attributes in UV space via CNNs, but drives 3D deformation through mesh-aware Jacobians, enabling smooth and semantically meaningful transitions across triangle boundaries. This hybrid design separates semantic modeling from geometric control, resulting in improved generalization, interpretability, and stability. Furthermore, TexAvatars captures fine-grained expression effects—including muscle-induced wrinkles, glabellar lines, and realistic mouth cavity geometry—with high fidelity. Our method achieves state-of-the-art performance under extreme pose and expression variations, demonstrating strong generalization in challenging head reenactment settings.
\end{abstract}

\section{Introduction}
% \label{sec:intro}
High-fidelity modeling of human head avatars remains a central challenge in graphics, especially with the rising demand for over 4K-quality avatars in telepresence, visual effects, and immersive media~\cite{pica}. Recent advances in 3D Gaussian Splatting (3DGS)~\cite{3dgs} have shown impressive results in photorealistic reconstruction and real-time rendering. Building on this, 3DGS-based head avatars generally follow two paradigms: mesh-driven deformation in 3D space or texel-driven attribute regression.  

Mesh-based avatars~\cite{gaussianavatars, surfhead, splattingavatar, rgba} leverage 3D Morphable Models (3DMMs)~\cite{flame, bfm} for analytic, triangle-wise rigging. This provides stability and interpretability, but linear mesh deformation struggles to capture fine non-linear expressions like wrinkles or asymmetric activations. Furthermore, since each Gaussian is tied to a single triangle, neighboring Gaussians may deform inconsistently across boundaries. ScaffoldAvatar~\cite{scaffoldavatar} recently addresses this by introducing spatial correlations in 3D, inspired by hierarchical frameworks~\cite{scaffoldgs}, enabling more coherent local control of facial regions.

Conversely, texel-based avatars~\cite{gem, rgca, gaussianheads, tega} leverage the continuous nature of UV space, where attributes are retrieved via interpolative $\mathrm{GridSample_\text{lerp}}$. Unlike analytic rigging methods that operate on individually bound Gaussians, this formulation introduces spatial correlation across nearby texels for smoother representations. However, most of these methods discard or minimally use 3DMM geometry and instead use CNNs to regress canonical-to-deformed offsets without explicit mesh-driven modeling. As a result, deformation is fully learned rather than analytically grounded, leading to poor extrapolation under extreme expressions (pose). Many also rely on heuristic offset regularizers, which are difficult to tune reliably.

We observe that mesh- and texel-based approaches offer complementary strengths, and our method is designed to leverage both. Mesh-based models inherently follow the mesh deformation, which provides coarse but, physically plausible initialization. Their use of local-to-global designs~\cite{gaussianavatars,surfhead,rgba} enables regularization at the normalized local level, making them well-suited for imposing physically meaningful constraints such as scale bounds or spatial locality in normalized frame. These formulations also allow Gaussians to deform coherently in accordance with the surface geometry. On the other hand, texel-based approaches benefit from operating in a CNN-regressed continuous UV space with a uniform sampling grid. This formulation naturally imposes spatial correlation among neighboring attributes, which introduces a structured inductive bias into learning~\cite{dip}. As a result, it facilitates better generalization and robustness in deformation prediction, especially when learned attributes are shared across similar spatial contexts. 

We propose a hybrid texel–3D rigging strategy that unifies the advantages of mesh-driven and texel-driven paradigms. Rather than predicting canonical-to-deformed offsets or directly regressing in the deformed space, we regress local Gaussian attributes in UV space. A key observation is that such attributes are \textbf{not spatially coherent until transformed into global texel space}: standard $\mathrm{GridSample_\text{lerp}}$ implicitly assumes local linearity, yet local positions and scales lack correlation in UV coordinates where adjacent texels may correspond to distant mesh regions. To resolve this, we remap triangle-driven Jacobians into texel space and apply deformation analytically prior to sampling, enabling interpolation to occur in a consistent, linearizable domain, called \emph{Quasi-Phong Jacobian Field}. This design preserves geometric fidelity while removing the rigid binding between Gaussians and mesh triangles~\cite{gaussianavatars,surfhead}, effectively binding them to a smoothly blended neighborhood. Finally, to compensate for the limitations of 3DMM-based expressions in capturing fine-scale dynamics, we introduce a lightweight latent expression embedding~\cite{emoportraits}, allowing opacity and color to reflect wrinkles, folds, and occlusions without the overhead of an additional shading branch.

In summary, our contributions are threefold. First, \textbf{Local Flexible Gaussians in Texel Space}. Instead of binding Gaussians to specific mesh triangles, we regress expression-dependent local attributes directly in texel space using CNNs. This design allows Gaussians to adaptively vary within their local coordinates, while still anchored by analytic rigging for robust generalization. Second, \textbf{Texel Space Coupling for Coherent Deformation.} By remapping triangle-wise deformations into texel space and applying them to CNN-predicted local attributes, we form a continuous Gaussian field across the surface. This deformation acts like a structured kernel over the CNN output—injecting mesh-aware gradients that stabilize training and reinforce geometric consistency. Last, \textbf{Superior Generalization.} Our approach outperforms prior methods by a significant margin, particularly in cross-identity and cross-expression driving, achieving robust rendering quality.
\section{Related Work}

\subsection{Human Head Modeling} 3D head modeling forms the foundation for dynamic avatar construction. Traditional 3D Morphable Models (3DMMs)~\cite{bfm} parameterize facial shape and appearance using PCA, but struggle to capture articulated components such as the jaw, neck, and eyeballs. To address this, FLAME~\cite{flame} integrates linear blend skinning (LBS) from body models like SMPL~\cite{smpl}, while still relying on PCA for expressions. More recently, NPHM~\cite{nphm, mononphm} proposed a neural SDF-based model that improves expressiveness and flexibility beyond FLAME, removing fixed topology. With the rise of neural radiance fields (NeRFs)~\cite{nerf}, methods such as NerFACE~\cite{nerface}, CAFCA~\cite{cafca}, and IMAvatar~\cite{imavatar} model dynamic heads using 3DMM-conditioned volumetric fields and blendshape-aware skinning. Other work~\cite{insta, rignerf, localdeformhead} leverage mesh-driven deformation, while neural rendering~\cite{diffusionavatars, nha, flare} enhances realism and modularity. Concurrently, UV-space modeling~\cite{mvp, monoavatars} enables convolution-friendly, continuous representations. 

\subsection{3DMM Rigging-based Head Avatars}

With the advent of 3D Gaussian Splatting (3DGS)\cite{3dgs}, the community has shifted toward fast, rasterization-friendly forward deformations. PointAvatar\cite{pointavatar} pioneered point-based avatars with isotropic splats driven by blendshape bases, extending IMAvatar into a splatting-friendly regime. Later works~\cite{gaussianavatars, splattingavatar, surfhead} tied Gaussians to 3DMM meshes via analytic deformations, avoiding MLPs to preserve real-time efficiency. Fully neural fields~\cite{gha, headgas, npga, flashavatar, rig3dgs} regress offsets with MLPs, but incur higher costs and weaker generalization. Hybrids~\cite{gbs, rgba} blend 3DMM rigging with neural offsets, yet still deform Gaussians independently, often lacking spatial coherence. ScaffoldAvatar~\cite{scaffoldavatar} alleviates this by patch-based correlation with anatomical priors. In contrast, our method leverages UV-unwrapped geometry and CNNs to couple Gaussians across texel space while retaining analytic mesh rigging. Unlike mesh-bound Gaussians~\cite{gaussianavatars,surfhead}, our attributes vary with expressions but remain anchored in continuous UV coordinates—yielding both flexible expressiveness and stable rigging.

\subsection{Texel-Based Head Avatars}

Texel-space methods exploit UV-unwrapped facial meshes for structured convolutions in dense attribute regression~\cite{pica, mega, gem, rgca, gaussianheads, gghead}.  
\citet{flashavatar} initializes Gaussians in UV space but applies no further deformation, while \citet{mvp} introduces analytic triangle-based rigging with UV-aligned TBN spaces.  
\citet{gem} regresses most attributes purely neurally, limiting semantic grounding and robustness under expression extremes.  
\citet{rgba} combines blendshape-based deformation with texel-space features, and \citet{rgca, mega} emphasize improved appearance decoding.  
\citet{gghead} follows a 3D GAN pipeline, predicting Gaussian attributes via UV-space CNNs.  
Despite these advances, most approaches rely on neural deformation with heuristic regularization or canonical-to-deformed mappings, which can be unstable in extreme or novel settings.  

\noindent\textbf{Concurrent Work.} TeGA~\cite{tega} maps canonical UVD Gaussians to 3D with Adaptive Density Control (ADC) and extra MLPs for deformation and shading.  
In contrast, we regress Gaussians per frame directly in UV space and lift them analytically to 3D, avoiding ADC and learned offsets, leading to better generalization under expression extrapolation.

\begin{figure}[t]
  \centering
  \includegraphics[width=0.9\linewidth]{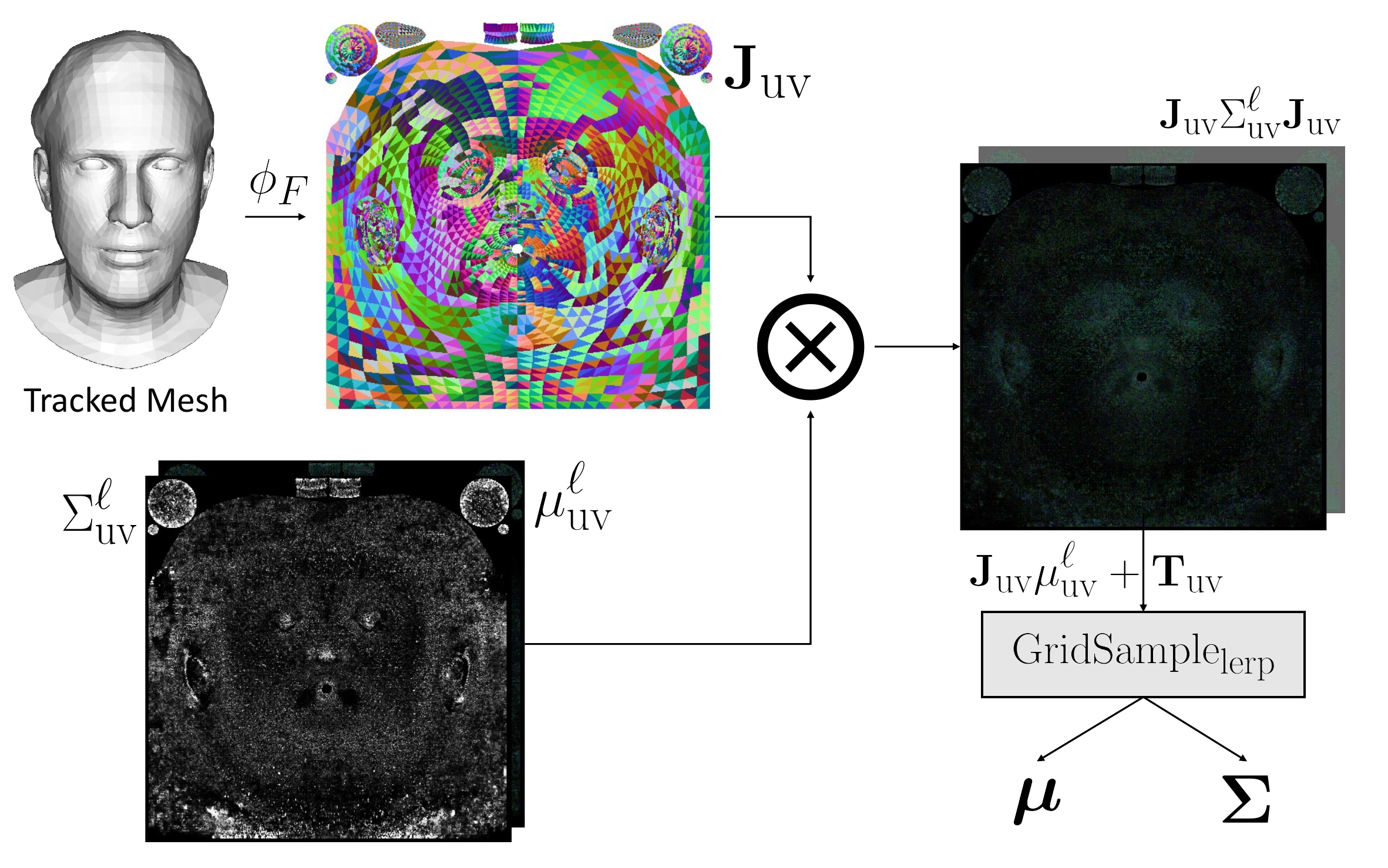}
  \vspace{-0.2cm}
    \caption{\textbf{Overview.} These local attributes are lifted to global space by transforming them with precomputed Jacobians \(\mathbf{J}_\text{uv}\), remapped from the tracked mesh to texel space by remapping function $\phi_F$. This results in globally coherent attributes that are continuous across surface regions via linear grid sampling.
}
  % \Description{Ablation figure to describe the effect of latent expression condition.}
  \label{fig:overview}
  \vspace{-0.4cm}
\end{figure}
\section{Preliminaries}
\subsection{3D Gaussian Splatting (3DGS)}

3D Gaussian Splatting (3DGS)~\cite{3dgs} represents a scene as a set of Gaussian primitives, each defined by a position $\mu \in \mathbb{R}^3$, rotation $r \in \mathbb{R}^4$, scale $s \in \mathbb{R}^3$, SH color $\mathbf{SH} \in \mathbb{R}^{3 \times 16}$, and opacity $\alpha \in \mathbb{R}$. The rotation and scale are combined into a covariance matrix $\Sigma = R S S^\top R^\top$, and rendering is performed by a tile-based differentiable rasterizer using camera parameters.

\subsection{Analytic Rigging-based Gaussian Head Avatars} 
GaussianAvatars~\cite{gaussianavatars} introduces an analytic rigging method that maps local Gaussians to global 3D space via \textbf{binding inheritance}, where each Gaussian is associated with a specific mesh triangle. During deformation, each triangle propagates its transformation—using isotropic scaling, barycentric positioning, and TBN-based rotation—to its bound Gaussians. SurFhead~\cite{surfhead} extends this idea by adopting a Jacobian-based transformation inspired by the DTF~\cite{dtf} formulation, replacing the scaled rotation model in GaussianAvatars. This enables modeling of both stretching and anisotropic scaling. Throughout this paper, we use superscript $\ell$ to denote local-space quantities. The global deformation is defined as:
\begin{equation}
    \boldsymbol{\Sigma} = \mathbf{J} \Sigma^{\ell} \mathbf{J}^\top, \quad 
    \mu = \mathbf{J} \mu^{\ell} + \mathbf{T},
\end{equation}
where $\mathbf{T}$ is the triangle centroid and $\mathbf{J}$ is the deformation matrix, realized as a scaled rotation in GaussianAvatars~\cite{gaussianavatars} and as a full Jacobian in SurFhead~\cite{surfhead}.

\subsection{3D Morphable Head Models (3DMMs) and Image Animation} 
3DMMs is a fundamental component in head avatar reconstruction. In our method, we assume access to per-frame tracked meshes, as obtained from a photometric tracker identical to that used in~\citet{gaussianavatars}. Each tracked mesh $M$ is parameterized by an identity-agnostic expression parameter $\psi \in \mathbb{R}^{100}$, an identity parameter $\beta \in \mathbb{R}^{300}$, and a rigid pose parameter $\theta \in \mathbb{R}^{15}$. Since our work focuses on single-identity fitting, we omit the identity parameter $\beta$ in the remainder of the formulation.

Image Animation, first introduced by~\citet{fsth}, transfers expression and pose from a driving image to a source image while preserving the source identity. In our setting, we extract an expression-related parameter $\eta \in \mathbb{R}^{128}$ from~\citet{emoportraits}, which serves as an auxiliary signal to the 3DMM expression code for transferring fine-grained expressions, such as wrinkles, that are not well captured by standard 3DMMs.
% \begin{figure*}[h]
%   \centering
%   \includegraphics[width=0.95\linewidth]{figures/fig_overview_final.pdf}  
%   \caption{\textbf{Overview.} Our pipeline begins with expression parameters \((\psi, \eta)\) and pose parameters \((\theta)\). A lightweight geometry decoder \(\mathcal{D}_{g}\) and appearance decoder \(\mathcal{D}_{a}\) predict local Gaussian attributes in texel space, including position \(\mu^{\ell}_\text{uv}\), rotation \(r^{\ell}_\text{uv}\), and scaling \(s^{\ell}_\text{uv}\). These local attributes are lifted to global space by transforming them with precomputed Jacobians \(\mathbf{J}_\text{uv}\), remapped from the tracked mesh to texel space. This results in globally coherent attributes that are continuous across surface regions via linear grid sampling in the tangent space. Additional attributes such as opacity \(\alpha_\text{uv}\) and color \(\mathbf{c}_\text{uv}\) are also regressed. Finally, the global attributes are passed to a tile-based Gaussian rasterizer \(\mathcal{R}\)~\cite{3dgs} to render high-fidelity images.
% }
% % \Description{Overall pipeline figure of our model.}
% \end{figure*}

\begin{figure*}[h]
  \centering
  \includegraphics[width=\linewidth]{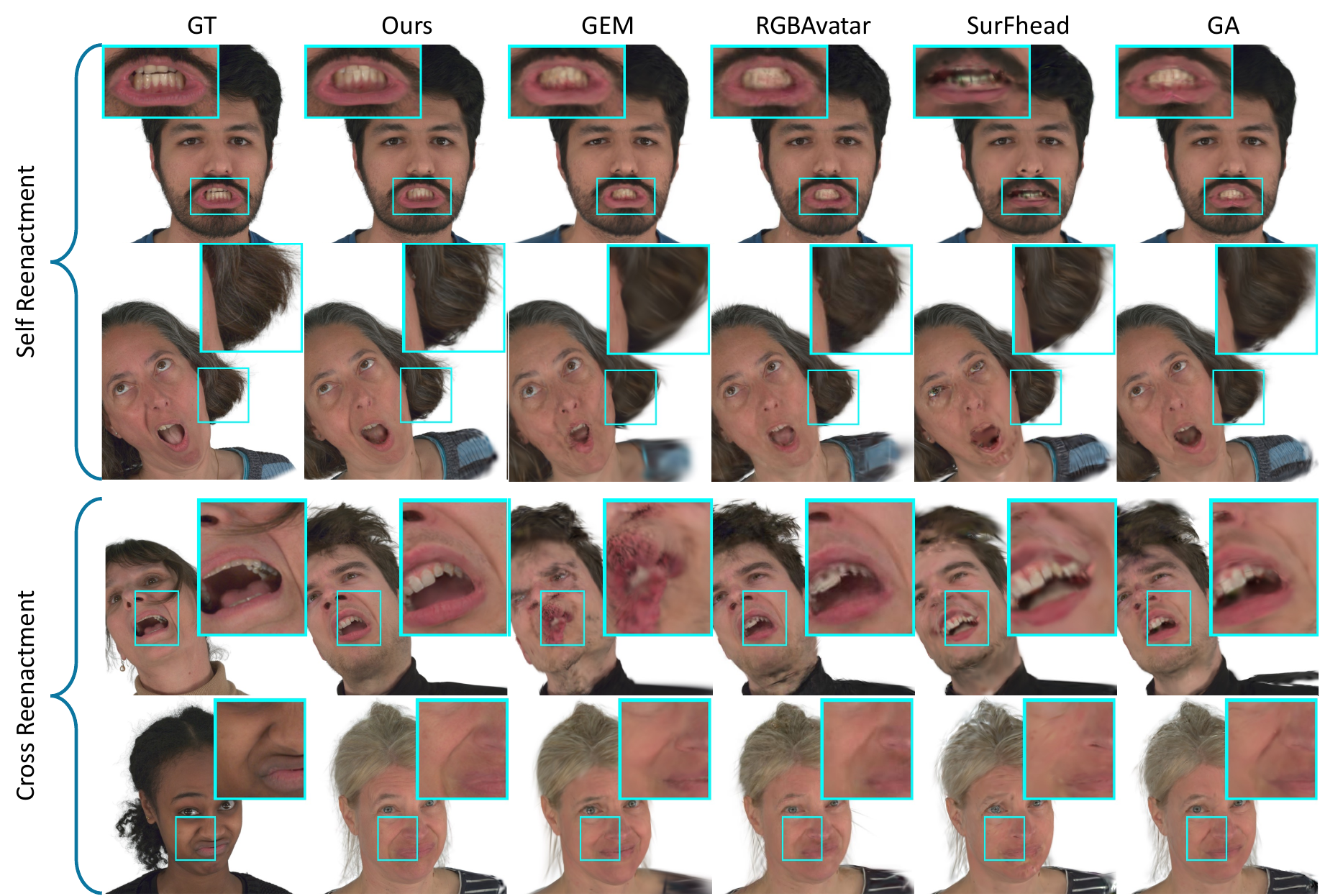}
  \vspace{-0.7cm}
  \caption{\textbf{Extreme Self- and Cross-Reenactment Scenario.} Our approach demonstrates significantly higher fidelity under extreme facial motions and rigid head rotations (subjects from NeRSemble~\cite{nersemble}, \textbf{FREE} corpus). Notably, our model accurately reconstructs high-frequency features such as nasolabial lines, hair strands, and detailed oral cavity structures. These results highlight the effectiveness of our hybrid texel-rigging framework in preserving semantic details even under highly expressive and challenging settings.
    }
    % \Description{Self-Reenactment comparison figure with extreme scenario.}
    \vspace{-0.4cm}
  \label{fig:quali_main}
\end{figure*}

\section{Method}

In this work, we introduce a novel parameterization that integrates the strengths of both mesh-based and UV-based rigging to achieve more robust and generalizable deformations. Instead of relying solely on fixed local Gaussians or black-box neural deformations, we regress local Gaussian attributes that are explicitly anchored to mesh triangles within the UV layout. These Gaussians are defined in a local texel space and subsequently mapped into a global (deformed)\footnote{In this work, since we perform local-to-global transformations without explicitly defining a canonical space, we use the term “global” to denote the deformed space.} texel space via triangle's transformations, similar to~\cite{gaussianavatars}. However, our formulation enables the model to exploit both the geometric structure of the mesh topology and the spatial continuity of texel-aligned CNN features. We begin in Sec.~\ref{method:dynamic_local} by describing how we regress local attributes, colors, and opacity using CNNs. In Sec.~\ref{method:rigging}, we explain how we construct a continuous global texel space using triangle-wise Jacobians derived from the 3DMM mesh. Finally, in Sec.~\ref{method:reg}, we present our regularization strategy, which operates directly in this continuous texel space.

\subsection{Local Texel-Attribute Parameterization}\label{method:dynamic_local}
\noindent\textbf{Intuition.}~\citet{gaussianavatars} and~\citet{surfhead} achieve efficient rendering by binding Gaussians to 3DMM meshes via analytic rigging from local to global space. However, in their setup, the local Gaussian attributes are fixed with respect to the expression parameters and merely follow the mesh deformation. This rigid coupling restricts expressiveness, as the Gaussians cannot adapt beyond the mesh articulation, leading to limited interpolation capacity. 

In contrast, we regress local Gaussian attributes with a network conditioned on expression, rather than fixing identical attributes across expressions, while still grounding them in analytic rigging. This yields twofold benefits: (i) the learned attributes increase representational flexibility within bounded local coordinates, enhancing interpolation; and (ii) since the predictions remain governed by the smooth global Jacobian, gradients are well-conditioned, acting as a stabilizing kernel. This prevents uncontrolled drift and supports more reliable extrapolation under out-of-distribution expressions.

\noindent\textbf{Analysis.} In the texel-based regime,~\citet{gem} and~\citet{rgca} predict canonical-to-deformed offsets or directly regressed deformed quantities. In their formulation, 
$G_d = G_c + \Delta G$, where $G$ denotes Gaussian attributes and subscripts $d$ and $c$ denote deformed and canonical respectively, the gradient of the image-space training loss $\mathcal{L}$ with respect to the Gaussian texel map decoder $f_\theta$'s parameters $\theta$ is
\[
\nabla_\theta \mathcal{L} \;=\; 
\frac{\partial \mathcal{L}}{\partial G_d}\;
\frac{\partial G_d}{\partial \theta} \;=\; 
\frac{\partial \mathcal{L}}{\partial G_d}\;
\frac{\partial \Delta G}{\partial \theta},
\]
which directly scales with the magnitude of global-space displacements $\Delta G$ and can therefore become unstable under large deformations. To mitigate this, prior methods often resort to heuristic regularizers that shrink offsets toward zero—e.g., penalizing scale or position displacements—to avoid divergence, at the cost of reduced expressiveness.

In contrast, our local parameterization adopts 
$G_d = \mathbf{T}(G_\ell)$ with $G_\ell = f_\theta(x)$ denoting local 
Gaussian attributes regressed in normalized local space. 
Consequently, $\tfrac{\partial G_\ell}{\partial \theta}$ remains bounded, 
independent of the global displacement scale. The gradient then becomes
\[
\nabla_\theta \mathcal{L} \;=\; 
\frac{\partial \mathcal{L}}{\partial G_d}\;
\underbrace{\frac{\partial G_d}{\partial G_\ell}}_{\mathbf{J}}\;
\frac{\partial G_\ell}{\partial \theta}.
\]
Here, $\mathbf{J}$ denotes the Jacobian of the analytic mesh-based 
transformation $\mathbf{T}$. Since face-level deformations are dominated 
by rotations and only mild shear or scaling, $\mathbf{J}$ is 
near-isometric with singular values close to unity, i.e., $\|\mathbf{J}\|\le C$. 
Thus the gradient magnitude is bounded:
\[
\|\nabla_\theta \mathcal{L}\| \;\le\; C\,
\Big\|\tfrac{\partial L}{\partial G_d}\Big\|\,
\Big\|\tfrac{\partial G_\ell}{\partial \theta}\Big\|,
\]
ensuring stable training, \textbf{even $G_{\ell}$ was predicted from neural networks}. In summary, because errors remain confined to 
bounded local coordinates and cannot be arbitrarily amplified in global 
space, our formulation balances interpolation capacity with extrapolation stability.  

\noindent\textbf{Design.} 
We employ two expression-dependent CNN decoders inspired by the architecture of~\citet{rgca}: 
a view-dependent appearance decoder \(\mathcal{D}_a\), and a view-independent geometry decoder \(\mathcal{D}_g\). 
Both decoders are conditioned on the FLAME expression parameters \(\psi\), pose parameters \(\theta\), 
and the image-driven expression code \(\eta\) from~\citet{emoportraits}. 
While \(\mathcal{D}_g\) predicts view-invariant geometry and opacity, 
\(\mathcal{D}_a\) additionally takes the view direction \(\pi\) as input to produce view-dependent appearance. 
We also replace the spherical-harmonics color representation with a precomputed RGB color 
$\mathbf{c} \in \mathbb{R}^3$, regressed per texel.

\begin{align}
G_{\text{uv}} &= \mathcal{D}_{g}(\psi, \theta, \eta), \quad
    \{\mu_{\text{uv}}^{\ell}, r_{\text{uv}}^{\ell}, s_{\text{uv}}^{\ell}, \alpha_{\text{uv}} \} \subset G_{\text{uv}} \\
A_{\text{uv}} &= \mathcal{D}_{a}(\psi, \theta, \eta, \pi), \quad
    \mathbf{c}_{\text{uv}} \in A_{\text{uv}}
\end{align}

Here, the subscript \(\text{uv}\) indicates texel-space attributes, while the superscript \(\ell\) denotes 
local-space representations. 
Unlike~\citet{tega}, who introduce an additional shading network to address baked-in effects such as wrinkles or ambient occlusion in Gaussian-based avatars, 
we find that simply predicting \emph{dynamic} color and opacity with the extra expression code $\eta$ is already sufficient to capture 
these appearance variations in practice, without requiring an extra shading stage.

\subsection{Rigging from Local Texel to 3D space}\label{method:rigging}
\paragraph{Na\"{i}ve Solution.}  
We begin with a naive approach where local Gaussian attributes---such as
rotation $r^{\ell}$, scale $s^{\ell}$, and position $\mu^{\ell}$---are
predicted in texel space and interpolated via bilinear sampling:
\begin{equation}
[\mu^{\ell}, r^{\ell}, s^{\ell}]
= \mathrm{GridSample}_{\mathrm{lerp}}([\mu^{\ell}_{\text{uv}}, r^{\ell}_{\text{uv}}, s^{\ell}_{\text{uv}}]).
\end{equation}
The interpolated attributes are then lifted to 3D space using the affine
transform of the associated triangle $F$:
\begin{equation}
\boldsymbol{\mu} = \mathbf{J}_F \mu^{\ell} + \mathbf{T}_F, \quad
\boldsymbol{\Sigma} = \mathbf{J}_F \Sigma^{\ell} \mathbf{J}_F^{\top}.
\end{equation}
Here, $\mathbf{J}_F$ and $\mathbf{T}_F$ denote the Jacobian and translation
for face $F$. While texels within the same triangle share a common local
frame, discontinuities arise across triangle boundaries, since each triangle
defines its frame independently. This leads to ambiguous or inconsistent
deformations when interpolating attributes across adjacent texels. Prior
binding strategies~\cite{gaussianavatars} inherit the same issue,
producing piecewise-discontinuous fields.

\begin{figure}[t]
  \centering
  \includegraphics[width=\linewidth]{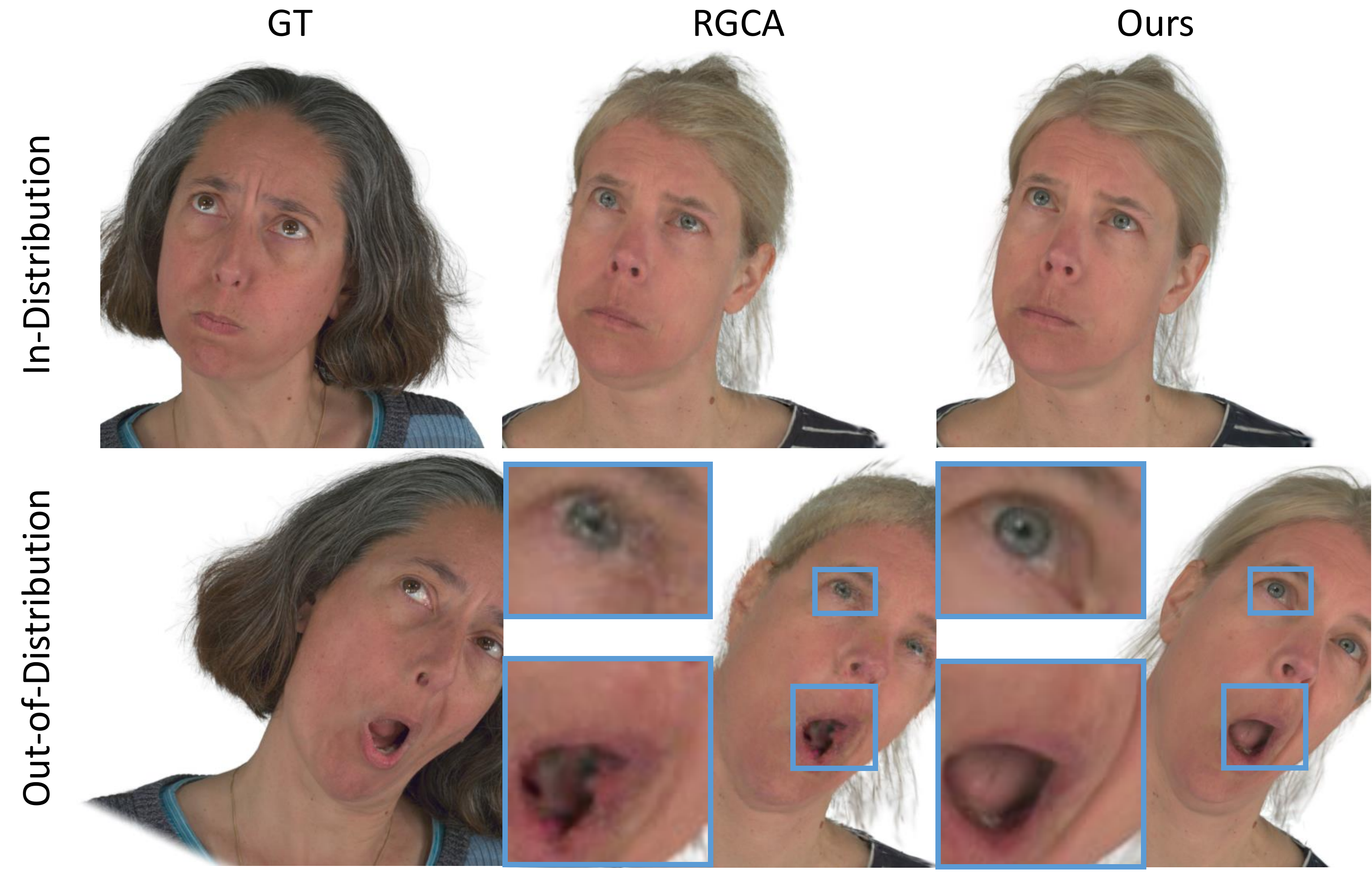}
  \vspace{-0.5cm}
  \caption{Comparison with RGCA~\cite{rgca}. RGCA adopts a relatively large mesh scale to stabilize training by keeping offsets small, which works well in most cases but can still cause blobby artifacts such as blurred nasal lines under strong stretching. }
  % \Description{Ablation figure to describe the effect of latent expression condition.}
  \vspace{-0.5cm}
  \label{fig:comp_rgca}
\end{figure}

\paragraph{Quasi-Phong Jacobian Field.}  
While our analytic rigging ensures stability, the per-face Jacobians 
$\mathbf{J}_F$ are inherently piecewise-constant across mesh faces, 
leading to discontinuities at face boundaries. To address this, we 
introduce a smooth global texel field by unwrapping triangle Jacobians 
into UV space.  
Let $\phi_F : \Delta^2 \to \Omega$ denote the parameterization map from 
the barycentric simplex $\Delta^2$ of face $F$ to the global UV domain 
$\Omega \subset \mathbb{R}^2$. We define the texel-space Jacobian field by
\begin{equation}
\mathbf{J}_{\mathrm{uv}}(u) \;=\; \sum_{F \in \mathcal{F}} 
\mathbf{1}_{\phi_F(\Delta^2)}(u)\;\mathbf{J}_F,
\end{equation}
where $\mathbf{1}_{\phi_F(\Delta^2)}$ indicates texels covered by face $F$.  

Although $\mathbf{J}_{\mathrm{uv}}$ is piecewise-constant, we resample it
via bilinear interpolation in UV space, yielding a continuous Jacobian field
analogous to Phong surface ~\cite{phongsurface}. We prioritize interpolation behavior, so we set \texttt{align\_corners=False} in $\mathrm{GridSample_{lerp}}$ \emph{even when the sampling size matches the source resolution}, ensuring center-based bilinear interpolation rather than corner-aligned identity mapping.
The final formulation is
\begin{equation}
\boldsymbol{\mu} = \mathrm{GridSample}_{\mathrm{lerp}}(\mathbf{J}_{\mathrm{uv}} \mu^{\ell}_{\mathrm{uv}} + \mathbf{T}_{\mathrm{uv}}),
\end{equation}
\begin{equation}
\boldsymbol{\Sigma} = \mathrm{GridSample}_{\mathrm{lerp}}(\mathbf{J}_{\mathrm{uv}} \Sigma^{\ell}_{\mathrm{uv}} \mathbf{J}_{\mathrm{uv}}^{\top}).
\end{equation}

\noindent Crucially, our network-predicted local scale and rotation parameters are merged into covariance form, ensuring that $\mathbf{\Sigma}$ still remains positive semi-definite. This guarantees that the resampled covariance lies in a smooth tangent space, enabling spatially coherent interpolation across texels. Fig.~\ref{fig:overview} provides an overview of our method.

Unlike~\citep{gaussianavatars}, which operate on piecewise flat surface frames, our formulation blends Jacobians across adjacent faces in UV space. This \emph{Quasi-Phong Jacobian Field} yields smooth, continuous deformations analogous to Phong shading, mitigating discontinuities and providing better-conditioned gradients for optimization.

\subsection{Constraint for Local Texel Space}\label{method:reg}
This step is crucial since our local space is not bound to a single triangle. 
Whereas~\citet{gaussianavatars} regularize local parameters within triangle coordinates, 
our texel-space Jacobian field allows interpolated texels to correspond to blended triangles, 
yielding a continuous family of global coordinate systems. 
Thus, local attributes only gain semantic meaning after Jacobian transformation, 
remaining valid even under blended interpolations. 
For stability, we follow~\citet{gaussianavatars} and impose per-texel lower bounds 
$\epsilon_\mu$ and $\epsilon_s$ on predicted positions and scales:
\begin{align}
\mathcal{L}_{\text{reg}_\mu} &= \|\max(\mu_\text{uv}, \epsilon_{\mu})\|, \\
\mathcal{L}_{\text{reg}_s} &= \|\max(s_\text{uv}, \epsilon_{s})\|.
\end{align}

\subsection{Optimization}
Our overall training objective integrates photometric reconstruction, perceptual similarity, and stability regularization into a unified loss:
\begin{equation}
\mathcal{L}_{\text{total}} = \mathcal{L}_{\text{recon}} + \lambda_{\text{vgg}} \cdot \mathcal{L}_{\text{VGG}} + \lambda_{\text{reg}_\mu} \cdot \mathcal{L}_{\text{reg}_\mu} + \lambda_{\text{reg}_s} \cdot 
 \mathcal{L}_{\text{reg}_s}.
\end{equation}
The reconstruction loss $\mathcal{L}_{\text{recon}}$ balances pixel-level accuracy and structural fidelity between the rendered image $I$ and the ground-truth $\hat{I}$:
\begin{equation}
\mathcal{L}_{\text{recon}} = \lambda_{L_1} \cdot \|I - \hat{I}\|_1 + \lambda_{\text{SSIM}} \cdot (1 - \text{SSIM}(I, \hat{I})).
\end{equation}
Following 3DGS~\cite{3dgs}, we set $\lambda_{L_1} = 0.8$ and $\lambda_{\text{SSIM}} = 1 - \lambda_{L_1}$. We incorporate a perceptual loss $\mathcal{L}_{\text{VGG}}$ computed using VGG features to enhance high-frequency fidelity. However, introducing it from the beginning destabilizes training and hampers PSNR and SSIM. To mitigate this, we activate the perceptual loss only after 300K iterations (out of 600K total), using a small weight $\lambda_{\text{vgg}}$ thereafter. As introduced in Sec.~\ref{method:reg}, we include two regularization terms to prevent degenerate Gaussians by enforcing lower bounds on predicted local means and scales. The coefficient $\lambda_{\text{reg}}$ controls their influence and remains active throughout training.

\definecolor{best}{RGB}{150, 200, 255}    % 진한 하늘색 (1등)
% \definecolor{second}{RGB}{190, 220, 255}  % 중간 하늘색 (2등)
\definecolor{second}{RGB}{230, 240, 255}   % 아주 연한 회청색 (3등)
\begin{table*}[t]
\centering
\vspace{-0.4 cm}
\caption{Simplified comparison across three evaluation settings: Novel Expression (\textbf{Held-out}), Novel Expression (\textbf{FREE}), and Novel View. \colorbox{best}{Best} and \colorbox{second}{second-best} scores are highlighted.}
\label{tab:grouped_metrics}
\resizebox{\textwidth}{!}{%
\begin{tabular}{l|ccc|ccc|ccc}
\toprule
& \multicolumn{3}{c|}{Novel Expression (\textbf{Held-out})} 
& \multicolumn{3}{c|}{Novel Expression (\textbf{FREE})} 
& \multicolumn{3}{c}{Novel View} \\
\cmidrule(r){2-4} \cmidrule(r){5-7} \cmidrule(r){8-10}
Method 
& LPIPS $\downarrow$ & SSIM $\uparrow$ & PSNR $\uparrow$ 
& LPIPS $\downarrow$ & SSIM $\uparrow$ & PSNR $\uparrow$ 
& LPIPS $\downarrow$ & SSIM $\uparrow$ & PSNR $\uparrow$ \\
\midrule
GaussianAvatars~\citep{gaussianavatars}
& 0.092 \ensuremath{\pm} 0.018 & \cellcolor{best} 0.897 \ensuremath{\pm} 0.028 & 25.00 \ensuremath{\pm} 2.84 
& 0.123 \ensuremath{\pm} 0.024 &  0.858 \ensuremath{\pm} 0.030 & 22.01 \ensuremath{\pm} 2.58 
& 0.087 \ensuremath{\pm} 0.021 & 0.919 \ensuremath{\pm} 0.021 &  29.23 \ensuremath{\pm} 1.82 \\
SurFhead~\cite{surfhead}
& 0.117 \ensuremath{\pm} 0.040 & 0.884 \ensuremath{\pm} 0.030 & 24.47 \ensuremath{\pm} 2.64  
& 0.150 \ensuremath{\pm} 0.036 & 0.854 \ensuremath{\pm} 0.035 & 22.06 \ensuremath{\pm} 2.53 
& 0.180 \ensuremath{\pm} 0.122 & 0.872 \ensuremath{\pm} 0.032 & 23.80 \ensuremath{\pm} 2.78 \\
RGBAvatar~\cite{rgba}
& 0.101 \ensuremath{\pm} 0.014 & 0.889 \ensuremath{\pm} 0.025 & 24.51 \ensuremath{\pm} 1.58 
& 0.134 \ensuremath{\pm} 0.019 & 0.860 \ensuremath{\pm} 0.028 & 21.76 \ensuremath{\pm} 1.68 
& 0.084 \ensuremath{\pm} 0.013 & 0.924 \ensuremath{\pm} 0.017 & 28.26 \ensuremath{\pm} 1.01 \\
GEM~\cite{gem}
& 0.199 \ensuremath{\pm} 0.025 & 0.885 \ensuremath{\pm} 0.036 & 24.12 \ensuremath{\pm} 1.82 
& 0.154 \ensuremath{\pm} 0.027 & \cellcolor{best}0.863 \ensuremath{\pm} 0.034 & 22.01 \ensuremath{\pm} 2.07
& 0.181 \ensuremath{\pm} 0.021 & 0.918 \ensuremath{\pm} 0.028 & 28.16 \ensuremath{\pm} 1.72 \\
RGCA~\cite{rgca}
& \cellcolor{second}0.050 \ensuremath{\pm} 0.013 & 0.890 \ensuremath{\pm} 0.032 & \cellcolor{second}25.55 \ensuremath{\pm} 2.07  
& \cellcolor{second}0.086 \ensuremath{\pm} 0.025 & 0.854 \ensuremath{\pm} 0.035 & \cellcolor{second}22.68 \ensuremath{\pm} 2.59
& \cellcolor{best}0.030 \ensuremath{\pm} 0.006 & \cellcolor{second}0.943 \ensuremath{\pm} 0.013 & \cellcolor{second}34.24 \ensuremath{\pm} 1.23 \\
\midrule
Ours
& \cellcolor{best} 0.048 \ensuremath{\pm} 0.013 & \cellcolor{second}0.894 \ensuremath{\pm} 0.030 & \cellcolor{best}25.61 \ensuremath{\pm} 2.10
& \cellcolor{best} 0.077 \ensuremath{\pm} 0.017 & \cellcolor{second}0.861 \ensuremath{\pm} 0.033 & \cellcolor{best}22.84 \ensuremath{\pm} 2.05
& \cellcolor{best} 0.030 \ensuremath{\pm} 0.005 & \cellcolor{best}0.947 \ensuremath{\pm} 0.013 & \cellcolor{best}35.15 \ensuremath{\pm} 1.33 \\
\bottomrule
\end{tabular}
}
\vspace{-0.4cm}
\end{table*}

\section{Experiment} 
\subsection{Baseline Nomination} We compare our method against five recent state-of-the-art baselines: GaussianAvatars~\cite{gaussianavatars}, SurFhead~\cite{surfhead}, RGBAvatar~\cite{rgba}, GEM~\cite{gem} and Relightable Gaussian Codec Avatar~\cite{rgca}.

\noindent\textbf{Analytic Rigging Methods}
\begin{itemize}
    \item GaussianAvatars (GA) employs explicit triangle-wise rigging using the TBN frame of 3DMM mesh triangles.
    \item SurFhead (SF) generalizes GA with a Jacobian-based deformation that supports stretching and anisotropic scaling, introducing Jacobian Blend Skinning by blending adjacent triangle Jacobians.
\end{itemize}

\noindent\textbf{Texel-Neural Hybrid Methods}
\begin{itemize}
    \item RGBAvatar (RGBA) formulates Gaussians in texel space with TBN-based binding, augmented by neural Gaussian blendshapes~\cite{gbs}.
    \item Gaussian Eigen Model (GEM) regresses canonical-to-deformed Gaussian offsets via a style-based CNNs~\cite{stylegan}.
    \item Relightable Gaussian Codec Avatar (RGCA) directly predicts deformed Gaussians (positions as mesh-relative offsets) and stabilizes training by enlarging mesh scale, keeping offsets and scales well-conditioned.
\end{itemize}

% \noindent Texel-neural hybrid methods are structurally similar to ours in that they use texel-aligned neural deformation, but differ in that they predict deformation directly in the deformed space, whereas we apply mesh-aware analytic transformation after regressing local texel attributes.

\subsection{Dataset} 
We utilize the NeRSemble~\cite{nersemble} dataset, following the same train, novel view and novel expression evaluation protocol as GaussianAvatars~\cite{gaussianavatars}. Specifically, training uses 10 corpora and 15 cameras, holding out 1 near-frontal camera for novel-view testing, while a single corpus is reserved for novel-expression evaluation. 
In addition, we introduce an extra test split, denoted as \textbf{FREE}, consisting of longer and more unconstrained sequences with arbitrary expressions and head motions, to assess generalization beyond scripted motions~\cite{nersemble}.

\subsection{Comparisons}\label{sec:comp}
As shown in Fig.~\ref{fig:quali_main}, GEM underperforms under extreme expressions and poses, such as strong neck rotations or fine-scale details like glabellar wrinkles and nasal lines, which are tightly linked to subtle muscle activations. While GEM performs reasonably on frontal or mildly expressive frames, it struggles with extreme scenarios due to its reliance on CNN-predicted deltas—from a fixed canonical space to the deformed space—for position, scale, rotation, and opacity. Although position is transformed via a scaled rotation Jacobian, scale and rotation are directly regressed, and color remains pose-invariant, often resulting in the loss of photometric richness. RA adopts a hybrid approach by combining analytic rigging with MLP-predicted blendshapes of Gaussian attributes. However, it still suffers under unseen expressions, as the learned blendshape basis does not extrapolate well—especially when expression parameters fall far from the training manifold. SF proposes a continuous deformation field via learned Jacobian blending weights. Yet, this data-driven blending can be unstable in OOD settings due to its reliance on local examples. GA, leveraging analytic rigging, demonstrates strong generalization across expressions and poses. However, it lacks fine-scale geometric expressiveness and fails to model anisotropic deformation due to its use of isotropic scaled rotation, leading to blob-like artifacts in curved regions—as highlighted in the zoomed-in areas of Fig.~\ref{fig:quali_main}.

In contrast, our method achieves consistently sharper reconstructions, accurately modeling fine wrinkles, realistic mouth cavities, and sharp boundaries around facial hair. We attribute this to our hybrid rigging mechanism, which combines analytic structure with the adaptability of neural regression. These qualitative improvements are corroborated by quantitative gains (Table~\ref{tab:grouped_metrics})

\noindent\textbf{Additional Comparison with RGCA~\cite{rgca}.} RGCA~\cite{rgca} was originally proposed as relightable head avatars. Their practical trick—scaling the tracked mesh relatively large so that offsets remain small—makes training stable and interpolation reliable, which is why we selected it as a strong baseline. However, the same design backfires: under strong stretching or out-of-distribution expressions, the tiny offset budget leads to dotted artifacts (Fig.~\ref{fig:comp_rgca}). In contrast, our method predicts Gaussians adaptively in normalized local UV space, achieving both stability and expressiveness without relying on such tricks.

\noindent\textbf{More Qualitatives.} Additional visual results, including extreme expressions and large head motions, are provided in the supplementary due to space constraints, further highlighting the robustness and fidelity of our approach.
% : our method achieves the best LPIPS and PSNR scores with huge margin across the challenging FREE test set. Notably, we outperform the next best baseline in PSNR by +5.92 on novel-view (35.15 vs. 29.23), +0.61 on held-out expressions (25.61 vs. 25.00), and +0.79 on FREE (22.85 vs. 22.06), while also showing reduced standard deviations—indicating stronger robustness and stability.

\subsection{Ablations} 
\begin{figure}[t]
  \centering
  \includegraphics[width=\linewidth]{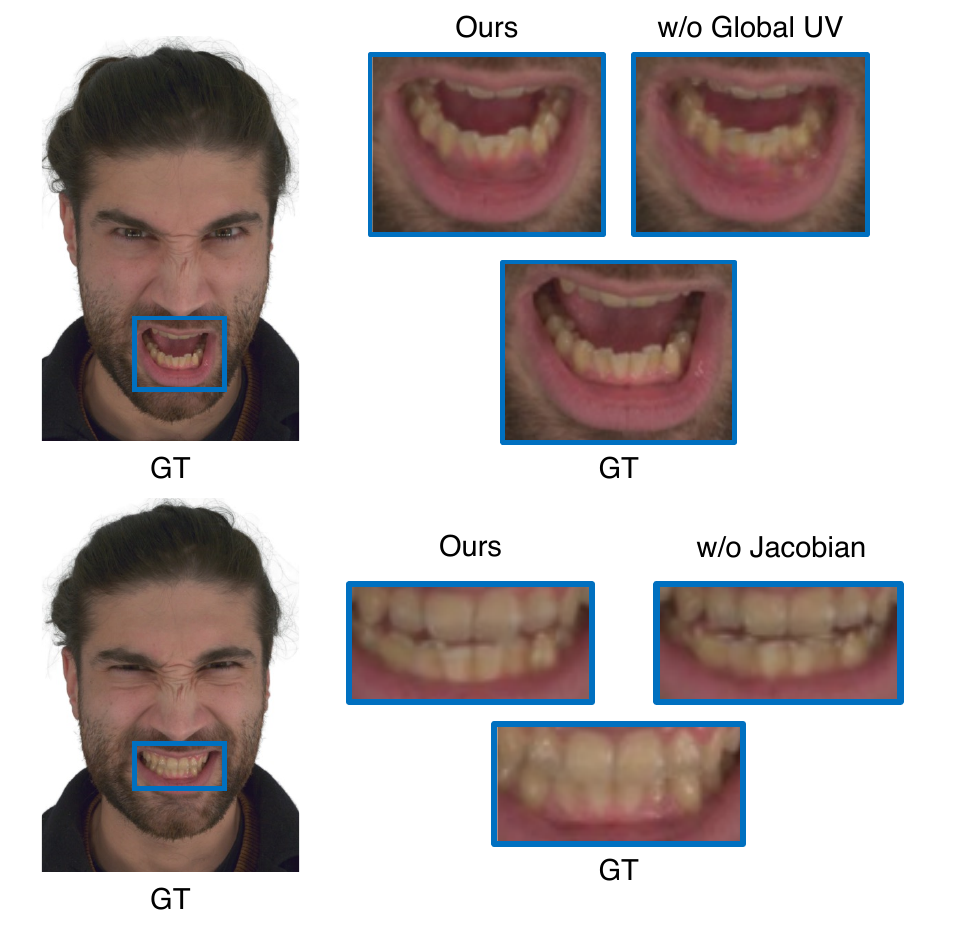}
  \vspace{-0.9cm}
  \caption{Effect of Global UV Sampling and Jacobian. Remapping mesh Jacobians to texel space enables smooth blending of attributes across triangle boundaries. Jacobian-based deformation effectively models stretch and anisotropic scaling while reducing blob-like artifacts.}
  % \Description{Ablation figure to describe the effect of global UV sampling and Jacobian transforms.}
  \label{fig:abl_globalUV}
    \vspace{-0.2cm}
\end{figure}

We curate four types of ablations including our contributions. Refer the Table~\ref{tab:ablation} for each ablation. Note that qualitative ablation for VGG loss is curated in supplementary.
\paragraph{Global vs. Local Grid Sampling}
Interpolating local Gaussian attributes (e.g., position and covariance) directly in local texel space leads to perceptual blurring, especially around regions with high geometric variation such as the mouth (Fig.~\ref{fig:abl_globalUV}). This occurs because these attributes are defined in triangle-specific local frames, and interpolation across triangle boundaries introduces semantic misalignment when UV-adjacent texels map to distant 3D positions. In contrast, we remap mesh-aware Jacobians into global UV space prior to interpolation, enabling geometrically coherent deformation blending. As shown in Fig.~\ref{fig:abl_globalUV} and supported by Table~\ref{tab:ablation}, our formulation better preserves detail and spatial coherence under challenging articulation.

\begin{figure}[t]
  \centering
  \includegraphics[width=0.9\linewidth]{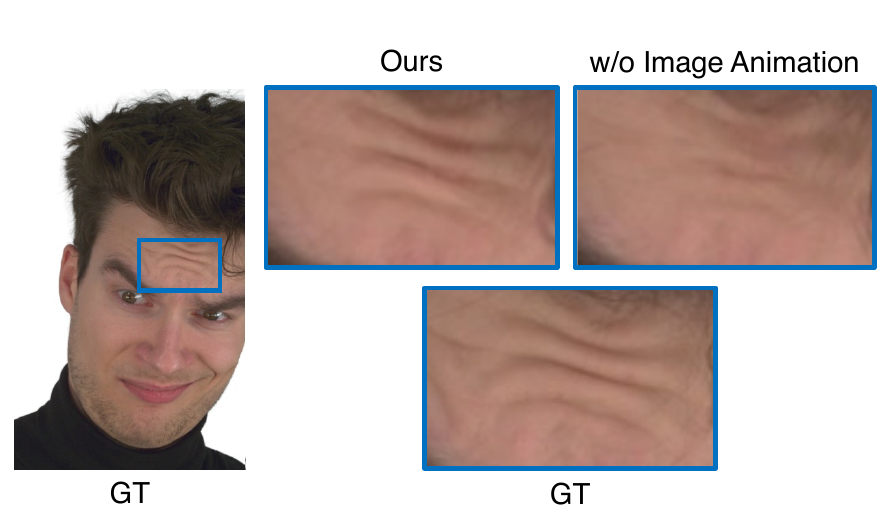}
  \vspace{-0.3cm}
  \caption{Effect of Image Animation Model. It enables the synthesis of details that are not explicitly represented in 3DMMs, such as wrinkles or subtle skin deformations, thereby enhancing realism under dynamic expressions. }
  % \Description{Ablation figure to describe the effect of latent expression condition.}
  \vspace{-0.3cm}
  \label{fig:abl_emo}
  
\end{figure}

\paragraph{Jacobian vs. Scaled Rotation} % 재성
Without stretch-friendly Jacobian, models relying solely on scaled rotation deformation—such as in GaussianAvatars~\cite{gaussianavatars}—struggle to capture anisotropic changes and are prone to producing blob-like artifacts in highly deformed regions. These limitations arise from the inability of isotropic scaling to account for directional stretch, leading to unnatural shape distortion.
\paragraph{Effect of Image Animation Model}  
 We find it crucial for capturing fine-grained details such as wrinkles. As shown in Fig.~\ref{fig:abl_emo}, even in self-driving scenarios, missing \(\eta\) can result in absent wrinkles, which we attribute to the limitations of the FLAME expression space. Since the same FLAME parameters can correspond to different surface details (e.g., with or without muscle activation), we introduce \(\eta\) to disambiguate such cases, and observe noticeable improvements.

\begin{table}
    % \vspace{-0.3cm}
  \caption{Ablation Studies for \textbf{FREE} testset. ‘w/o’ denotes removal of an individual component (not sequential subtraction); mean and standard deviation are computed across identities within a single run. \colorbox{best}{Best} score is highlighted.}
  \vspace{-0.1cm}
  \label{tab:ablation}
  \centering
    \resizebox{\linewidth}{!}{
  \begin{tabular}{l|ccc}
    \toprule
    Method & LPIPS~$\downarrow$ & SSIM~$\uparrow$ & PSNR~$\uparrow$ \\
    \midrule
    % Vector & 19.48&0.749&0.186 \\
    % \midrule
   Ours &  \cellcolor{best}0.077 \ensuremath{\pm} 0.017 &  \cellcolor{best} 0.861 \ensuremath{\pm} 0.033 &  \cellcolor{best} 22.84 \ensuremath{\pm} 2.05 \\
    \midrule
    w/o VGG       & 0.096 \ensuremath{\pm} 0.021 & 0.859 \ensuremath{\pm} 0.034 & 22.66 \ensuremath{\pm} 2.24 \\
    w/o Global UV & 0.096 \ensuremath{\pm} 0.028 & 0.855 \ensuremath{\pm} 0.035 & 22.46 \ensuremath{\pm} 2.39 \\
    % w/o Pl\"{u}cker & 19.95&0.774&0.166 \\
    w/o Jacobian & 0.080 \ensuremath{\pm} 0.018 &0.859 \ensuremath{\pm} 0.035 & 22.77 \ensuremath{\pm} 2.13 \\
    w/o Image Animation ($\eta$) & 0.078 \ensuremath{\pm} 0.017 & 0.853 \ensuremath{\pm} 0.034 & 22.69 \ensuremath{\pm} 1.97 \\
    \bottomrule
  \end{tabular}
    \vspace{-0.9cm}
  }
\end{table}
\section{Conclusion}
While TexAvatars demonstrates robust extrapolation of expression and pose across both self- and cross-driving scenarios, several limitations remain. Our pipeline relies on local texel space predictions modulated by a smooth Jacobian field via linear grid sampling in tangent space, which provides spatially coherent and analytically grounded rigging. This allows us to capture high-frequency details such as wrinkles and nasolabial folds, as well as complex cavities like the mouth interior. However, the method is not without its constraints. We provide detailed discussion in the supplementary material and briefly highlight key points: 1) \textbf{Hair Motion.} Our mesh-based representation cannot model dynamic hair, leading to artifacts under motion. Future work could incorporate strand-based dynamics~\cite{hhavatar}.
2) \textbf{Tongue Articulation.} FLAME~\cite{flame} does not include tongue geometry; thus, articulation inside the mouth is not captured. Extending the topology is a promising direction.
3) \textbf{Specular Effects.} High-frequency view-dependent effects like eye glints or sebum are insufficiently modeled; explicit specular rendering could address this.

\noindent 4) \textbf{Fixed Number of Primitives.} For the sake of training stability, the number of Gaussians is fixed throughout optimization. While this design choice sacrifices certain high-frequency details (e.g., pores and facial hair), it also opens up avenues for future work to better capture such fine-scale structures.

{
    \small
    \bibliographystyle{ieeenat_fullname}
    \bibliography{main}
}

\clearpage
\setcounter{page}{1}
\maketitlesupplementary

\newpage
\appendix

\section{Discussion on Future Work} 

\noindent\textbf{Hair motion.} In our experiments, we observed that elastic dynamics of hair—such as bouncing or swaying—are often present in the recorded dataset. Because our system is based on FLAME meshes, which lack of hair strands, such non-rigid and dynamic movements cannot be modeled within the current pipeline. Moreover, we noticed occasional entanglement between expression embeddings and hair motion, leading to jittering artifacts during cross-reenactment. This occurs due to the relatively high degrees of freedom in our parameterization (e.g., per-texel geometry and dynamic color/opacity). Incorporating dynamic hair representations, as explored in HHAvatar~\cite{hhavatar}, could be a fruitful direction, though it lies beyond the scope of this work.

\noindent\textbf{Tongue articulation.} Our approach depends heavily on the quality and completeness of the tracked 3DMM mesh~\cite{flame}. Since current FLAME-based meshes do not include tongue geometry, our model cannot reproduce tongue articulation, which is essential for expressive speech-driven animation. Future work may involve augmenting the 3DMM topology to include the tongue and other intraoral structures like~\citet{ltt_tracking}.

\noindent\textbf{Specular appearance.} Realistic rendering of specular effects remains challenging in our pipeline. Although view-dependent appearance modeling offers partial solutions, they are insufficient for fine-grained specularities, such as eye glints, tooth gloss, or sebum-induced facial shine. Extending our model to explicitly handle specular reflectance and integrate lighting control or relighting would be a valuable future direction.

\noindent\textbf{Fixed Number of Primitives.} Since we follow the convention of texel-based avatars~\cite{mvp, rgca, gem, gaussianheads}, which rely on uniform grid sampling in UV space, Gaussians are evenly distributed across the texels. This sometimes leads to missing high-frequency details such as pores or facial hair, which require a denser population of Gaussians. A quick remedy can be found in TeGA~\cite{tega}, which introduces learnable texel coordinates with 3DGS’s ADC. Exploring ways to allocate more Gaussians adaptively in high-frequency regions would be an intriguing research direction.

\begin{figure*}[h]
  \centering
  \includegraphics[width=\linewidth]{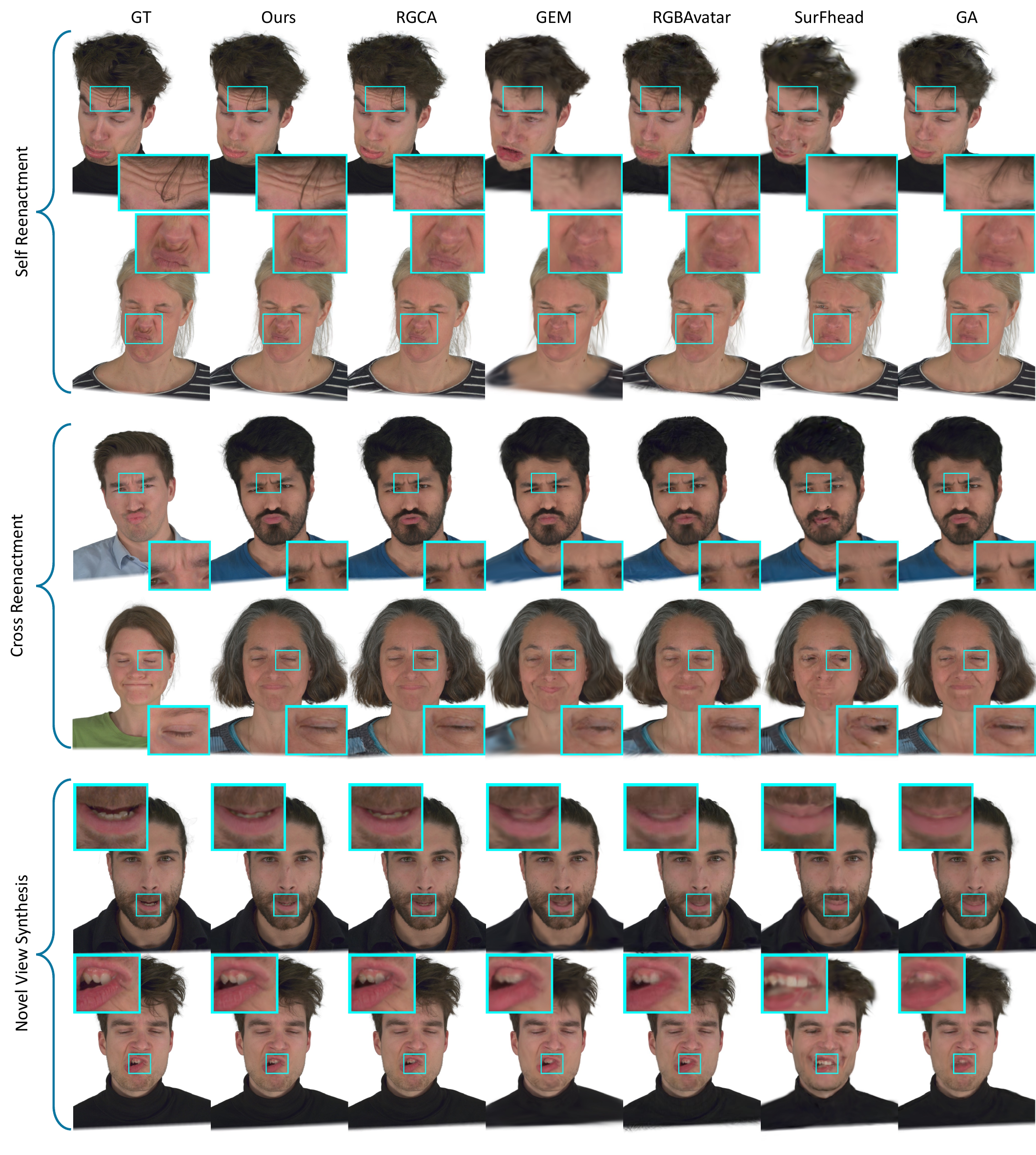}
  \caption{\textbf{Comparison Across Self-Reenactment, Cross-Reenactment, and Novel View Synthesis.} In comparison to RGCA, GEM, RGBAvatar, SurFhead and GaussianAvatars, our approach demonstrates significantly higher fidelity under extreme facial motions and rigid head rotations (subjects from NeRSemble~\cite{nersemble}, \textbf{FREE} corpus) and detailed reconstruction of frontal view (NeRSemble, \textbf{Validation} set including frontal views only). Notably, our model accurately reconstructs high-frequency features such as glabellar wrinkles, nasolabial lines, and detailed oral cavity structures. Fine-grained elements like eyebrow tension and hair strand separation are also more faithfully rendered. These results highlight the effectiveness of our hybrid texel-rigging framework in preserving semantic details even under highly expressive and challenging settings.
    }
    % \Description{Self-Reenactment comparison figure with extreme scenario.}
  \label{fig:quali_self_free}
\end{figure*}

\section{Ethical Considerations}

Our research presents a novel approach for reconstructing photorealistic 3D head avatars using 3DGS~\cite{3dgs} from multi-view images. While this method holds promise for advancing virtual communication and immersive digital experiences, the high fidelity of our reconstructions raises potential risks related to identity misuse, impersonation, and unauthorized reproduction of a person’s likeness. Since our approach currently requires a controlled multi-camera setup and data collection with a large span of expressions, such malicious use or unauthorized reenactment is much less likely to happen compared to other few-shot methods.

To better mitigate aforementioned issues, we only utilize NeRSemble~\cite{nersemble} data of participants who have provided explicit, signed consent for academic use. We also advocate for future work in secure model watermarking and identity verification to further safeguard digital likenesses. By proactively identifying and addressing these risks, we aim to ensure that the development and application of 3D neural avatars proceed in a socially responsible and ethically healthy direction.

\section{Additional Experiments}

\begin{figure}[t]
  \centering
  \includegraphics[width=\linewidth]{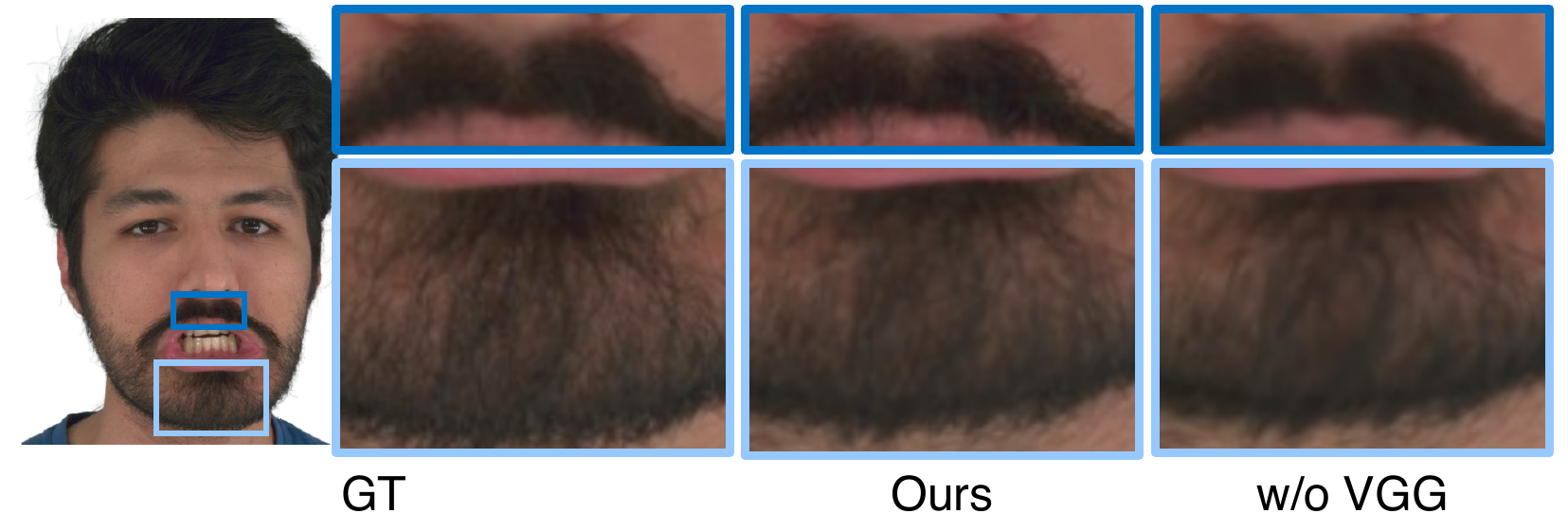}
  \caption{Effect of VGG Loss. Incorporating perceptual VGG loss enhances the reconstruction of fine-grained details, such as facial hair, by encouraging high-frequency fidelity.}
  % \Description{Ablation figure for effect of VGG loss.}
  \label{fig:abl_vgg}
\end{figure}

% \begin{figure*}[t] 
%   \centering
%   \includegraphics[width=\linewidth]{figures/Quali_val.pdf}
%   \caption{\textbf{Mild Self-Reenactment Scenario.} In comparison to GEM, RGBAvatar, SurFhead, and GaussianAvatars, our method achieves notably higher fidelity under extreme facial motions and rigid head rotations (subjects from NeRSemble~\cite{nersemble}, validation corpus). In particular, our model excels at reconstructing high-frequency details such as subtle wrinkles, fine structures around the teeth, and facial hair strands. These results underscore the strength of our hybrid texel-rigging framework in faithfully preserving semantically important microstructures, even in highly expressive and challenging scenarios.
%     }
%     % \Description{Self-Reenactment comparison figure with mild scenario.}
%   \label{fig:quali_val}
% \end{figure*}

\subsection{Ablation for Perceptual Loss}
 Following 3DGS~\cite{3dgs}, we adopt a combination of L1 and SSIM losses for image reconstruction. However, we observed that even when SSIM and PSNR scores are high, high-frequency details such as hair and beard often appear blurry. To better capture these fine details, we incorporate a perceptual loss based on VGG~\cite{vgg}. As noted in prior work such, Nerfies~\cite{nerfies}, which attributes the issue to gauge ambiguity---where PSNR tends to favor overly smooth or blurred outputs. As illustrated in Figure~\ref{fig:abl_vgg}, adding perceptual loss significantly improves fine-scale detail, which is not fully captured by pixel-level metrics.

\subsection{Qualitative Results on Validation Set.}
While our primary focus is on extreme cross-reenactment scenarios, we also report results on the novel-view validation set from NeRSemble~\cite{nersemble}. As shown in Fig.~\ref{fig:quali_self_free}, our method effectively reconstructs high-frequency details such as subtle wrinkles, fine structures around the teeth, and facial hair strands. While RGCA appears competitive in terms of facial details, its design is optimized for self-reenactment scenarios and thus struggles in real-world applications where novel-view and extreme-pose combinations prevail, as discussed in Section~\ref{sec:comp}. These results demonstrate the robustness of our hybrid texel-rigging framework, even in novel-view settings, and its ability to preserve semantically meaningful microstructures under various expressions and poses.

\section{Training Details}
\subsection{Hardware Information}
We trained our model for 600K iterations using a single batch on an NVIDIA RTX 3090 Ti GPU, which took approximately 16 hours. For FPS evaluation, we adopted the benchmark protocol from~\cite{gaussianavatars}, and our method achieved a real-time frame rate of 50.85 FPS. Notably, our approach is both lightweight and efficient: training typically requires only 6--10 GB of GPU memory, making it feasible to run on commonly available hardware (e.g., a single NVIDIA RTX 2080 Ti GPU with 12 GB VRAM).

Compared to GEM~\cite{gem}’s UNet baseline, which reported 33.90 FPS, our method demonstrates significantly superior runtime performance. Furthermore, our architecture is compatible with GEM's distillation pipeline, indicating that further improvements via distillation techniques remain a promising direction.

The learning rates were configured as follows: $lr_{g} = lr_{a} = 0.0006$ for the geometry/appearance decoders $\mathcal{D}_{g}$ and $\mathcal{D}_{a}$, and $lr_{\text{exp2code}} = lr_{\text{view2code}} = 0.0004$ for the expression/view encoders $\mathcal{F}_{\text{exp}}$ and $\mathcal{F}_{\text{view}}$ (see Fig.~\ref{fig:detailedlayers}). We used the Adam optimizer~\cite{adam} with the specified learning rates and $\epsilon = 1\text{e}{-15}$.

% Fig.~\ref{fig:abl_globalUV}
\begin{figure}[t]
  \centering
  \includegraphics[width=\linewidth]{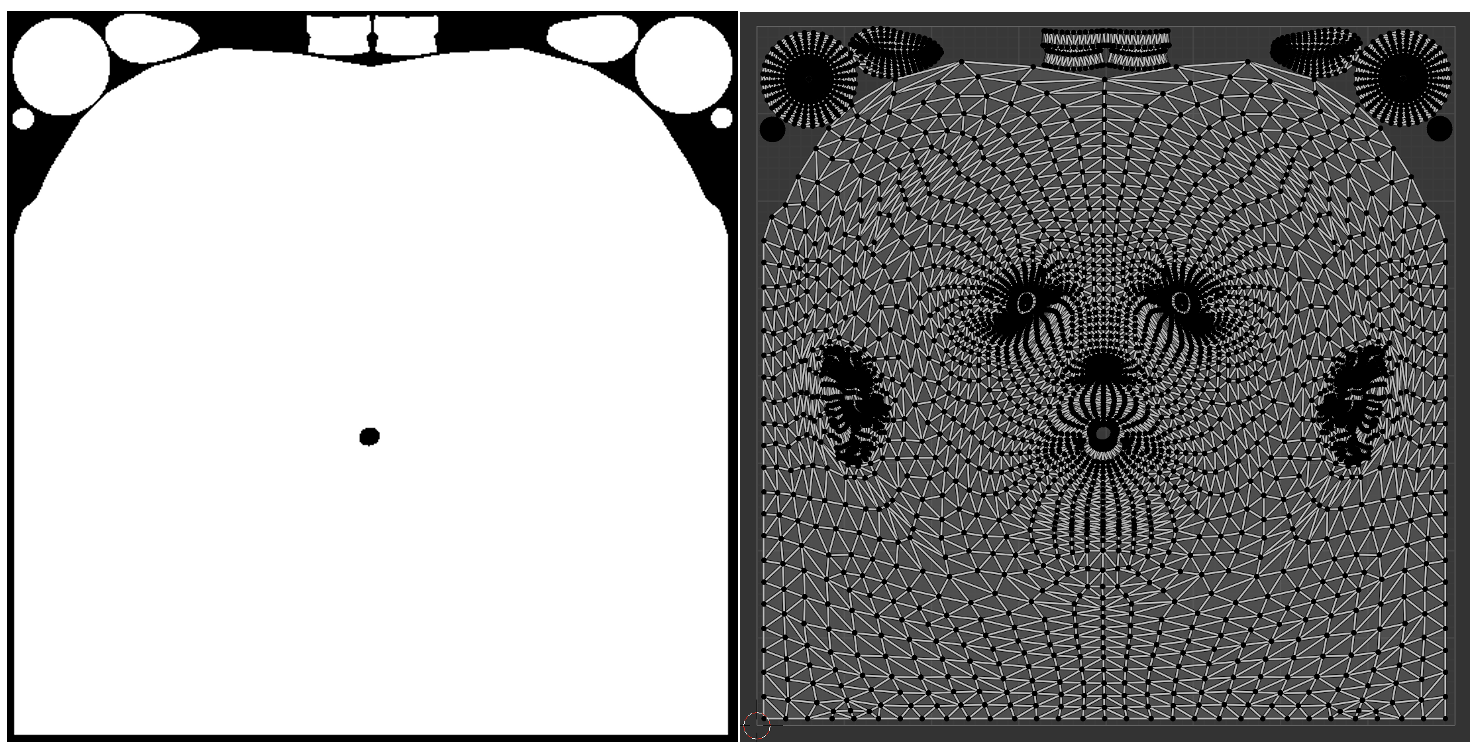}
  \caption{UV layout visualization. (Left) Binary mask showing UV coverage. White regions indicate valid UV coordinates covered by at least one triangle, while black regions denote invalid areas with no coverage. (Right) The corresponding unwrapped UV mesh map with triangulation overlay.}
  % \Description{Visualization of the occupied texel space in our unwrapped UV map and the detailed topology.}
  \label{fig:uvmap}
\end{figure}

\subsection{UV Map}

We utilize Blender 4.3.2~\cite{blender} to construct a custom UV layout based on the 2023 version of the FLAME model, which is identical to the base template adopted in GaussianAvatars ~\cite{gaussianavatars}. While GaussianAvatars programmatically incorporates 120 additional vertices with 168 faces for the upper and lower teeth, we explicitly include these structures in the UV layout, allowing the neural decoder to more effectively model the intraoral region. Furthermore, we introduce a tongue mesh consisting of 318 vertices and 632 faces, which is independently designed and integrated into our pipeline. The tongue is spatially divided into upper and lower parts and strategically placed between the teeth and eyeball regions within the UV space. The teeth and tongue UV faces are constructed by first applying `Minimum Stretch' UV-unwrapping function to the mesh then hand-crafted dilation, rotation and translation so that the faces occupy the map as much as possible (Fig. ~\ref{fig:uvmap}).

Since our Gaussian representation is rigged to the mesh and UVs outside valid regions are disregarded during the grid sampling stage, the addition of the tongue mesh does not sacrifice the allocation of Gaussians of other facial areas. In particular, when the tongue is visible in the training data, our mesh design helps avoid misplacing Gaussians on the teeth when the tongue is present, which often happens in models without an explicit tongue. This leads to better separation of oral structures and improves the realism of cavity rendering.

% \begin{figure}[h]
%   \centering
%   \includegraphics[width=\linewidth]{figures/UV.png}
%   \caption{UV layout visualization. (Left) Binary mask showing UV coverage. White regions indicate valid UV coordinates covered by at least one triangle, while black regions denote invalid areas with no coverage. (Right) The corresponding unwrapped UV mesh map with triangulation overlay.}
%   \Description{How unwrap our own UV map topology}
%   \label{fig:uvmap}
% \end{figure})

\subsection{Model Architecture}

\begin{figure*}[h]
  \centering
  \includegraphics[width=0.8\linewidth]{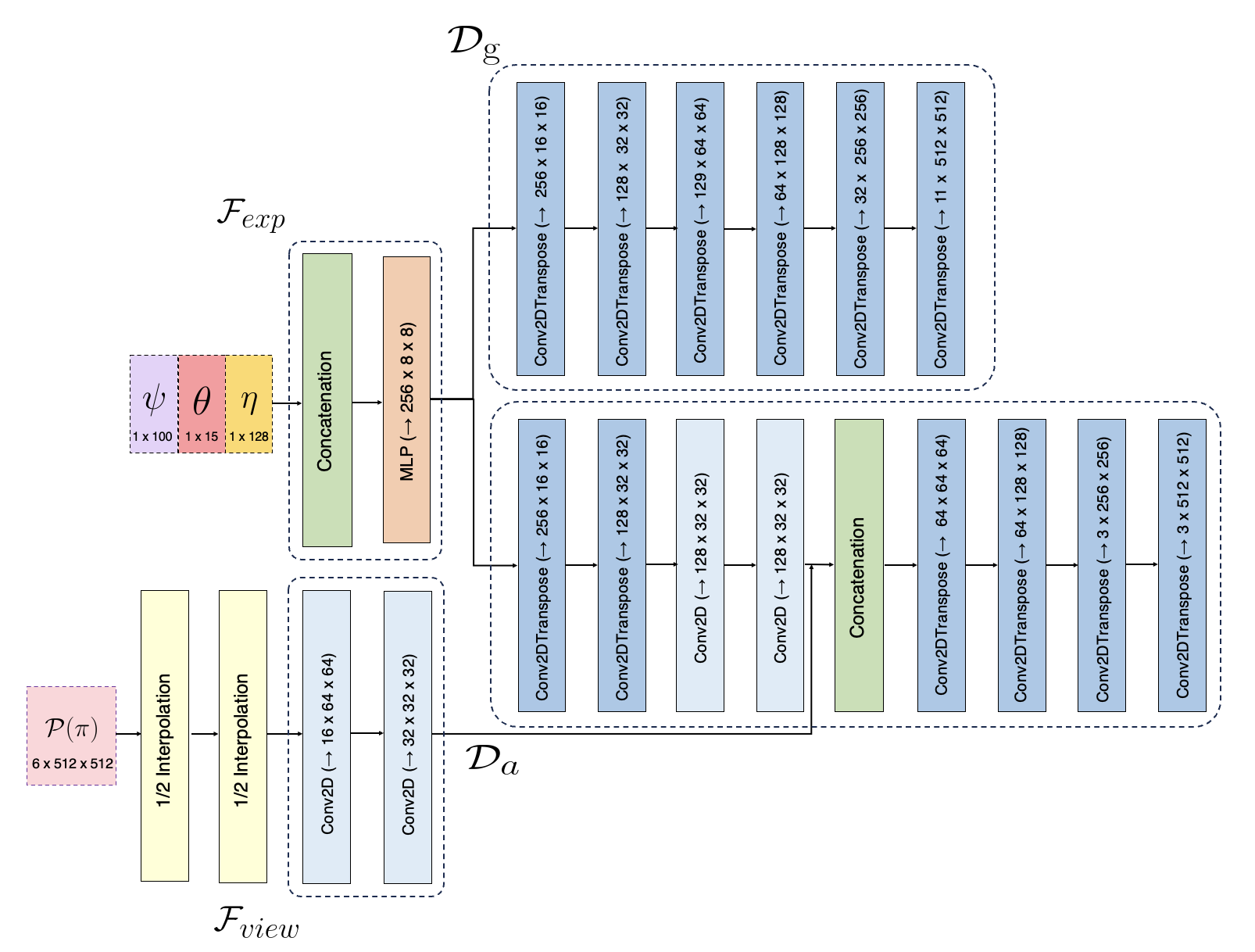}
  \caption{Detailed architecture of TexAvatars. Note that LeakyReLU (0.2) is applied after every layer except the final output layer (omitted from the figure for clarity).}
  % \Description{Detailed architecture of the network, mainly about our geometry and appearance decoders.}
  \label{fig:detailedlayers}
\end{figure*}

To enable real-time rendering of expressive 3D avatar representations, we propose a relatively shallow convolutional neural decoder architecture based on 2D CNNs and LeakyReLU activations (Fig. ~\ref{fig:detailedlayers}). Our design is inspired by the efficient structure of RGCA ~\cite{rgca}, modifying it to better suit the targeted regression of Gaussian-based appearance and geometry maps.

The model takes as input a driving signal composed of FLAME~\cite{flame} expression parameters ($\psi$), pose parameters ($\theta$), and EMOPortraits~\cite{emoportraits}  embeddings ($\eta$). These three components are stacked along the channel dimension and passed through a lightweight MLP, resulting in a compact expression code of dimension 256 $\times$ 8 $\times$ 8. This embedding is then separately processed by two decoders: a geometry decoder $\mathcal{D}_g$ and an appearance decoder $\mathcal{D}_a$.

Each decoder follows a cascade of 2D transposed convolutions with LeakyReLU activation (slope = 0.2), gradually upsampling the feature maps to a final resolution of 512 $\times$ 512. The geometry decoder outputs an 11-channel map representing the Gaussian offsets, while the appearance decoder directly produces a 3-channel RGB map. All convolutional and MLP layers are weight-normalized, following the original RGCA design for stable training and better convergence.

To provide viewpoint-aware conditioning, we compute a Plücker ray map ~\cite{plucker} from the known camera matrices of the NeRSemble ~\cite{nersemble} dataset. These maps, initially sized at the training resolution of 550 $\times$ 802, are resized to 512 $\times$ 512 and downsampled through two bilinear interpolation layers and two stride-2 2D convolutions, producing a 32 $\times$ 32 spatial feature embedding. This Plücker embedding is concatenated (along the channel dimension) with the intermediate feature map in $\mathcal{D}_a$ after its second LeakyReLU activation.

Based on our observations, directly injecting Plücker information at the final 512 $\times$ 512 resolution in order to preserve the semantic meanings rather tends to introduce significant color distortions across different viewpoints. While increasing the model capacity through deeper layers could potentially address this issue, such an approach would compromise the real-time performance of our system. In contrast, the proposed early-stage fusion at the 32 $\times$ 32 resolution allows the network to effectively leverage view-dependent cues while preserving both visual consistency and computational efficiency.

\subsection{Data Curation and Expression Coverage}

Training expressive, generalizable 3D avatars requires not only photometrically rich datasets but also a structured coverage of facial articulation. To this end, we curate a targeted subset of the NeRSemble dataset following \citet{gaussianavatars}, which offers high-resolution, multi-view recordings of spontaneous and posed facial behaviors. While the dataset originally includes a wide variety of motion corpora—including head gestures and occluded sequences—we focus on 14 semantically labeled corpora that emphasize emotionally and anatomically meaningful expressions (e.g., “Laugh,” “Cheeks,” “Mouth”). Note they are categorized into four emotion and six expression sequences, where one sequence for each subject is randomly held out as a self-reenactment evaluation set.

\begin{figure*}[h]
  \centering
  \includegraphics[width=1.0\linewidth]{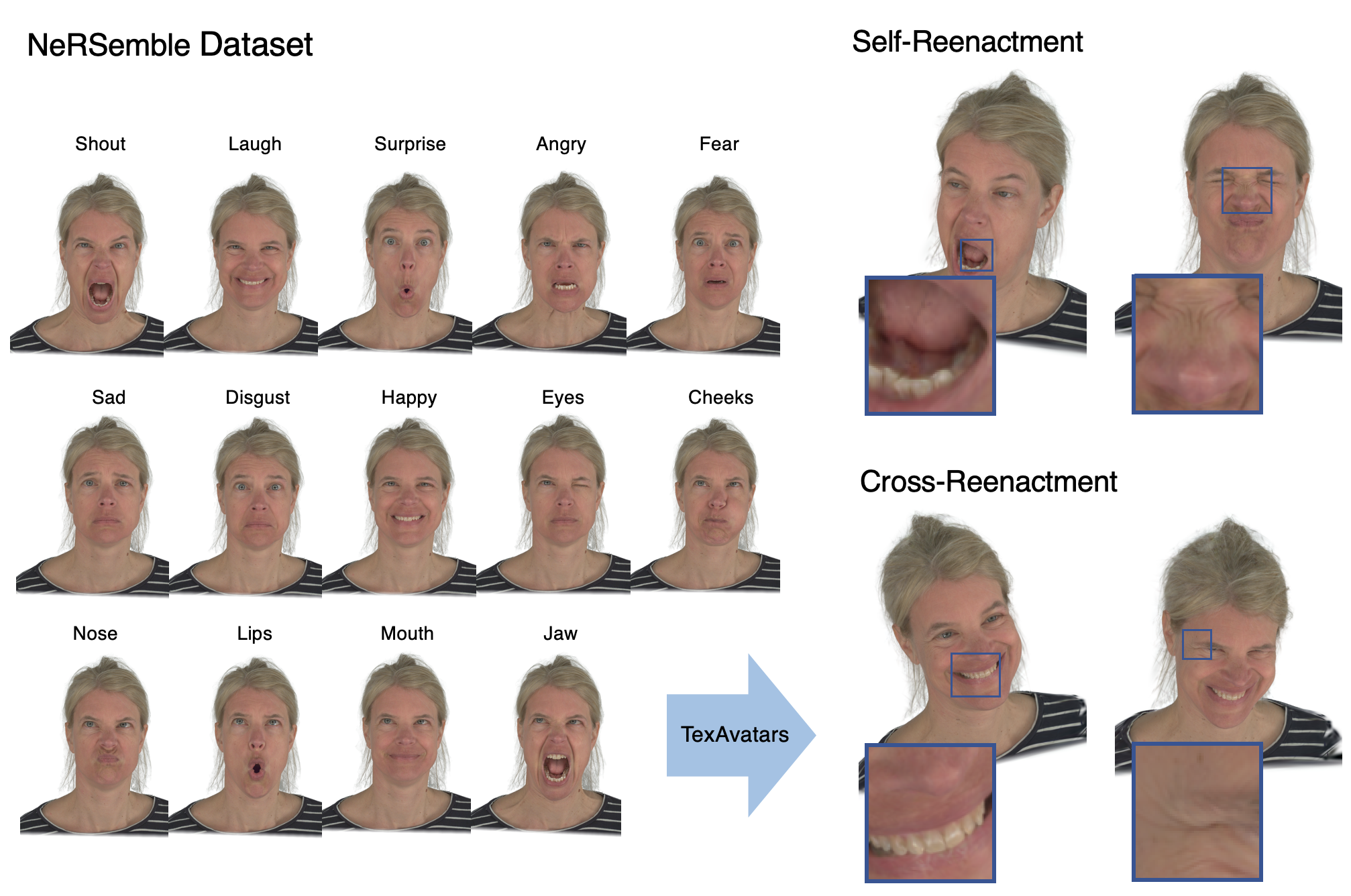}
  \caption{Overview of the NeRsemble dataset and reenactment results using TexAvatars. \textit{Left:} A diverse set of expressions and local facial changes from the NeRSemble dataset, including global emotions (e.g., Shout, Laugh, Fear) and localized actions (e.g., Eyes, Jaw, Mouth). \textit{Right:} Qualitative results of self-reenactment and cross-reenactment using TexAvatars.}
  % \Description{Visualization of the capability of TexAvatars to render images with dynamic expression and pose combinations far from NeRSemble dataset distribution}
  \label{fig:nersemble_curation}
\end{figure*}

To ensure both expression diversity and view consistency, 15-out-of-16 camera training split has been adopted. Expression corpora are drawn from ten unique motion sequences, each capturing variations in facial actuation. Unlike strictly scripted datasets, NeRSemble dataset exhibits natural variability within each label, ranging from subtle smirks to exaggerated shouts—allowing us to probe how well a model generalizes beyond rigid definitions of emotion as curated in Fig.~\ref{fig:nersemble_curation} \textit{(Left)}.

Our method leverages this curated set to learn neural avatars capable of reenactment across both self and cross identities. In self-reenactment, although subjects tend to revisit a narrow band of familiar expressions, our model faithfully reconstructs high-frequency details—such as the gentle folding around the nose and mouth, and faint shapes of the teeth as shown in Fig.~\ref{fig:nersemble_curation} \textit{(Right)}.

Cross-reenactment, by contrast, presents a more challenging task that expressions from one identity are imposed on a different subject’s geometry and texture space. Despite the distributional gap, TexAvatars preserves delicate expression-dependent features, including eye squints, tongue-teeth separation, and fine wrinkles around dynamic regions including the mouth and cheeks (Fig.~\ref{fig:nersemble_curation} \textit{(Right)}).

Interestingly, although tongue sequences are excluded from training, TexAvatars demonstrates a capacity to infer plausible tongue geometry from the limited scenes. This suggests that our model does not merely interpolate within the data manifold, but instead learns a consistent, anatomy-informed prior that can generalize to rarely or entirely unseen expressions with minimal supervision.

\end{document}